\documentclass[a4paper,11pt]{article}
\usepackage{amsmath}
\usepackage{amsmath,amssymb,epsfig}

\newcommand{\EQ}{\begin{equation}}
\newcommand{\EN}{\end{equation}}
\newcommand{\bea}{\begin{eqnarray}}
\newcommand{\eea}{\end{eqnarray}}

\input{tcilatex}

\begin{document}

\title{A conformal field theory description of magnetic flux
fractionalization in Josephson junction ladders}
\author{Gerardo Cristofano, Vincenzo Marotta\thanks{
Work supported in part by the European Communities Human Potential Program
under contract HPRN-CT-2000-00131 Quantum Spacetime} \\
%EndAName
Dipartimento di Scienze Fisiche, Universit\'{a} di Napoli ``Federico II''
and INFN, Sezione di Napoli - Via Cintia - Compl. universitario M.
Sant'Angelo - 80126 Napoli, Italy \\
E-mail: gerardo.cristofano@na.infn.it, vincenzo.marotta@na.infn.it \and %
Adele Naddeo \\
%EndAName
Dipartimento di Scienze Fisiche, Universit\'{a} di Napoli ``Federico II''
and Coherentia-INFM, Unit\`{a} di Napoli - Via Cintia - Compl.universitario
M. Sant'Angelo - 80126 Napoli, Italy \\
E-mail: adele.naddeo@na.infn.it \and Giuliano Niccoli \\
%EndAName
Sissa and INFN, Sezione di Trieste - Via Beirut 1 - 34100 Trieste, Italy \\
E-mail: niccoli@sissa.it}

%--------------------------------------------------------------------------

%--------------------------------------------------------------------------
%--------------------------------------------------------------------------

\thispagestyle{empty}\rightline{Napoli DSF-T-05/2004}%
\rightline{INFN-NA-05/2004}\rightline{ISAS/24/2004/FM} \vspace*{2cm}

\begin{center}
{\LARGE A conformal field theory description of magnetic flux
fractionalization in Josephson junction ladders}

{\LARGE \ }

{\large Gerardo Cristofano\footnote{{\large {\footnotesize Dipartimento di
Scienze Fisiche,}\textit{\ {\footnotesize Universit\'{a} di Napoli
\textquotedblleft Federico II\textquotedblright\ \newline
and INFN, Sezione di Napoli}-}{\small Via Cintia - Compl.\ universitario M.
Sant'Angelo - 80126 Napoli, Italy}}}, Vincenzo Marotta\footnotemark[1]  , }

{\large Adele Naddeo\footnote{{\large {\footnotesize Dipartimento di Scienze
Fisiche,}\textit{\ {\footnotesize Universit\'{a} di Napoli ``Federico II'' 
\newline
and Coherentia-INFM, Unit\`{a} di Napoli}-}{\small Via Cintia - Compl.\
universitario M. Sant'Angelo - 80126 Napoli, Italy}}}, Giuliano Niccoli%
\footnote{{\large \textit{\footnotesize Sissa and INFN, Sezione di Trieste -
Via Beirut 1 - 34100 Trieste, Italy}}} }

{\small \ }

\textbf{Abstract\\[0pt]
}
\end{center}

\begin{quotation}
We show how the recently proposed effective theory for a Quantum Hall system
at \textquotedblleft paired states\textquotedblright\ filling $\nu =1$ \cite%
{cgm2}\cite{cgm4}, the twisted model (TM), well adapts to describe the
phenomenology of Josephson Junction ladders (JJL) in the presence of
defects. In particular it is shown how naturally the phenomenon of flux
fractionalization takes place in such a description and its relation with
the discrete symmetries present in the TM. Furthermore we focus on
\textquotedblleft closed\textquotedblright\ geometries, which enable us to
analyze the topological properties of the ground state of the system in
relation to the presence of half flux quanta.

\vspace*{0.5cm}



{\footnotesize Work supported in part by the European Communities Human
Potential}

{\footnotesize Program under contract HPRN-CT-2000-00131 Quantum
Spacetime\newpage }\baselineskip=18pt \setcounter{page}{2}
\end{quotation}

\section{Introduction}

Arrays of weakly coupled Josephson junctions provide an experimental
realization of the two dimensional ($2D$) XY model physics. A Josephson
junction ladder (JJL) is the simplest quasi-one dimensional version of an
array in a magnetic field \cite{ladder}; recently such a system has been the
subject of many investigations because of its possibility to display
different transitions as a function of the magnetic field, temperature,
disorder, quantum fluctuations and dissipation. In this paper we focus on
the phenomenon of fractionalization of the flux quantum $\frac{hc}{2e}$ in a
fully frustrated JJL, the basic question being how the phenomenon of Cooper
pair condensation can cope with properties of charge (flux)
fractionalization, typical of a low dimensional system with a discrete $%
Z_{2} $ symmetry.

We must recall that charge fractionalization has been successfully
hypothesized by R. Laughlin to describe the ground state of a strongly
correlated $2D$ electron system, a quantum Hall fluid, at fractional
fillings $\nu =\frac{1}{2p+1}$, $p=1,2,...$. In such a system charged
excitations are present with fractional charge (anyons) and elementary flux $%
\frac{hc}{e}$. Furthermore the phenomenon of fractionalization of the
elementary flux has been found in the description of a quantum Hall fluid at
non standard fillings $\nu =\frac{m}{mp+2}$ \cite{cgm2}\cite{cgm4}, within
the context of $2D$ Conformal Field Theories (CFT) with a $Z_{m}$ twist.

In Refs. \cite{noi1}\cite{noi2} it has been shown that the presence of a $%
Z_{2}$ symmetry accounts for more general boundary conditions for the
propagating electron fields which arise in quantum Hall systems in the
presence of impurities or defects. Furthermore such a symmetry is present
also in the fully frustrated XY (FFXY) model or equivalently, see Ref. \cite%
{foda}\cite{noi}, in two dimensional Josephson junction arrays (JJA) with
half flux quantum $\frac{1}{2}\frac{hc}{2e}$ threading each square cell and
accounts for the degeneracy of the ground state.

It is interesting to notice that it is possible to generate non trivial
topologies, i.e. the torus, in the context of a CFT approach. That allows in
our case to show how non trivial global properties of the ground state wave
function emerge and how closely they are related to the presence of half
flux quanta, which can be viewed also as \textquotedblleft topological
defects\textquotedblright .

The aim of this paper is to show that the twisted model (TM) well adapts to
describe the phenomenology of fully frustrated JJL with a topological defect
and to analyze the implications of \textquotedblleft
closed\textquotedblright\ geometries on the ground state global properties.

The paper is organized as follows:

In Section 2 we introduce the physics of a fully frustrated JJL evidencing
the underlying $Z_{2}$ symmetry and then present the modified ladder with a
topological defect.

In Section 3 we describe the role played by such a symmetry in the
construction of the TM model and its relation with the ladder physics.
Furthermore the degeneracy of the ground state appears to be closely related
to the number of excitations (primary fields) of the CFT description.

In Section 4 the symmetry properties of the ground state conformal blocks
are analyzed and its relation with their topological properties shown.

In Section 5 a brief summary of the results is presented together with some
comments and suggestions.

In the Appendix the TM conformal blocks are explicitly given in terms of its
boundary states (BS) content \cite{noi1}\cite{noi2}.

\section{Josephson junction ladder with a topological defect}

In this Section, after describing the general properties of a ladder of
Josephson junctions as drawn in Fig.1, we introduce an interaction of the
charges (Cooper pairs) with a magnetic impurity (defect), as drawn in Fig.
2. With each site $i$ we associate a phase $\varphi _{i}$ and a charge $%
2en_{i}$, representing a superconducting grain coupled to its neighbors by
Josephson couplings; $n_{i}$ and $\varphi _{i}$ are conjugate variables
satisfying the usual phase-number commutation relation. The Hamiltonian
describing the system is given by the quantum phase model (QPM): 
\begin{equation}
H=-\frac{E_{C}}{2}\sum_{i}\left( \frac{\partial }{\partial \varphi _{i}}%
\right) ^{2}-\sum_{\left\langle ij\right\rangle }E_{ij}\cos \left( \varphi
_{i}-\varphi _{j}-A_{ij}\right) ,  \label{act0}
\end{equation}
where $E_{C}=\frac{\left( 2e\right) ^{2}}{C}$ ($C$ being the capacitance) is
the charging energy at site $i$, while the second term is the Josephson
coupling energy between sites $i$ and $j$ and the sum is over nearest
neighbors. $A_{ij}=\frac{2\pi }{\Phi _{0}}$ $\int_{i}^{j}A{\cdot }dl$ is the
line integral of the vector potential associated to an external magnetic
field $B$ and $\Phi _{0}=\frac{hc}{2e}$ is the magnetic flux quantum. The
gauge invariant sum around a plaquette is $\sum_{p}A_{ij}=2\pi f$ with $f=%
\frac{\Phi }{\Phi _{0}}$, where $\Phi $ is the flux threading each plaquette
of the ladder. Let us label the phase fields on the two legs with $\varphi
_{i}^{\left( a\right) }$, $a=1,2$ and assume $E_{ij}=E_{x}$ for horizontal
links and $E_{ij}=E_{y}$ for vertical ones. Let us also make the gauge
choice $A_{ij}=+\pi f$ for the upper links, $A_{ij}=-\pi f$ for the lower
ones and $A_{ij}=0$ for the vertical ones, which corresponds to a vector
potential parallel to the ladder and taking opposite values on upper and
lower branches.

Thus the effective quantum Hamiltonian (\ref{act0}) can be written as \cite%
{ladder}: 
\begin{eqnarray}
-H &=&\frac{E_{C}}{2}\sum_{i}\left[ \left( \frac{\partial }{\partial \varphi
_{i}^{\left( 1\right) }}\right) ^{2}+\left( \frac{\partial }{\partial
\varphi _{i}^{\left( 2\right) }}\right) ^{2}\right] +  \notag \\
&&\sum_{i}\left[ E_{x}\sum_{a=1,2}\cos \left( \varphi _{i+1}^{\left(
a\right) }-\varphi _{i}^{\left( a\right) }+\left( -1\right) ^{a}\pi f\right)
+E_{y}\cos \left( \varphi _{i}^{\left( 1\right) }-\varphi _{i}^{\left(
2\right) }\right) \right] .  \label{ha1}
\end{eqnarray}

\begin{figure}[tbp]
\centering\includegraphics*[width=0.7\linewidth]{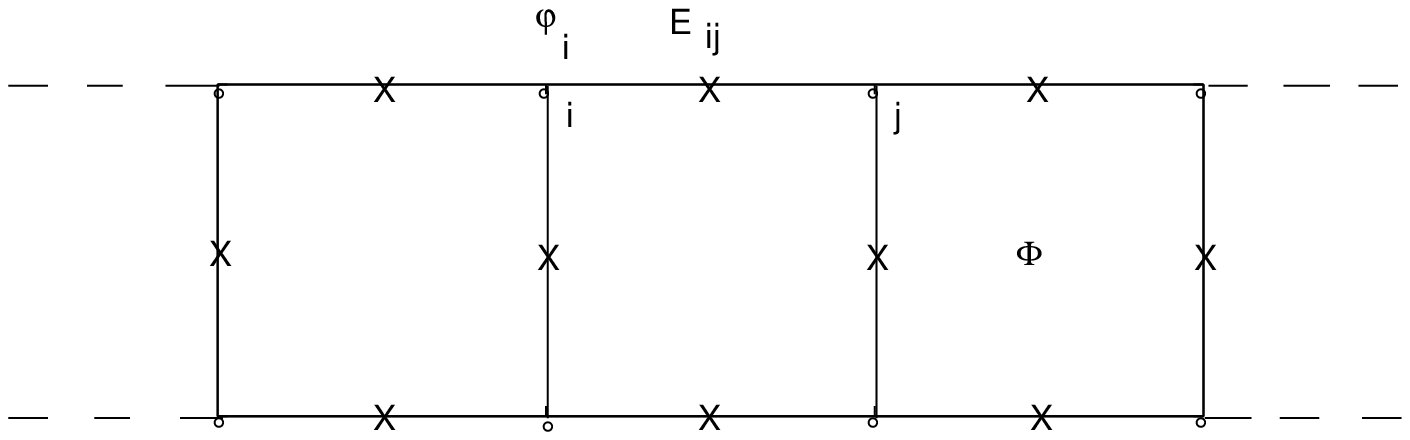}
\caption{Josephson junction ladder }
\label{figura1}
\end{figure}

The correspondence between such Hamiltonian and our TM model can be best
shown performing the change of variables: $\varphi _{i}^{\left( 1\right)
}=X_{i}+\phi _{i}$, $\varphi _{i}^{\left( 2\right) }=X_{i}-\phi _{i}$, so
eq. (\ref{ha1}) can be cast in the form: 
\begin{eqnarray}
-H &=&\frac{E_{C}}{2}\sum_{i}\left[ \left( \frac{\partial }{\partial X_{i}}%
\right) ^{2}+\left( \frac{\partial }{\partial \phi _{i}}\right) ^{2}\right] +
\notag \\
&&\sum_{i}\left[ 2E_{x}\cos \left( X_{i+1}-X_{i}\right) \cos \left( \phi
_{i+1}-\phi _{i}-\pi f\right) +E_{y}\cos \left( 2\phi _{i}\right) \right] ,
\label{ha2}
\end{eqnarray}
where $X_{i}$, $\phi _{i}$ (i.e. $\varphi _{i}^{\left( 1\right) }$, $\varphi
_{i}^{\left( 2\right) }$) are only phase deviations of each order parameter
from the commensurate phase and should not be identified with the phases of
the superconducting grains \cite{ladder}.

When $f=\frac{1}{2}$ and $E_{C}=0$ (classical limit) the ground state of the 
$1D$ frustrated quantum XY (FQXY) model displays - in addition to the
continuous $U(1)$ symmetry of the phase variables - a discrete $Z_{2}$
symmetry associated with an antiferromagnetic pattern of plaquette
chiralities $\chi _{p}=\pm 1$, measuring the two opposite directions of the
supercurrent circulating in each plaquette. Thus it has two symmetric,
energy degenerate, ground states characterized by currents circulating in
the opposite directions in alternating plaquettes. For small $E_{C}$ there
is a gap for creation of kinks in the antiferromagnetic pattern of $\chi
_{p} $ and the ground state has quasi long range chiral order.

Performing the continuum limit of the Hamiltonian (\ref{ha2}): 
\begin{eqnarray}
-H &=&\frac{E_{C}}{2}\int dx\left[ \left( \frac{\partial }{\partial X}%
\right) ^{2}+\left( \frac{\partial }{\partial \phi }\right) ^{2}\right] + 
\notag \\
&&\int dx\left[ E_{x}\left( \frac{\partial X}{\partial x}\right)
^{2}+E_{x}\left( \frac{\partial \phi }{\partial x}-\frac{\pi }{2}\right)
^{2}+E_{y}\cos \left( 2\phi \right) \right]  \label{ha3}
\end{eqnarray}%
we see that the $X$ and $\phi $ fields are decoupled. In fact the $X$ term
of the above Hamiltonian is that of a free quantum field theory while the $%
\phi $ one coincides with the quantum sine-Gordon model. Through an
imaginary-time path-integral formulation of such a model \cite{zinn} it can
be shown that the $1D$ quantum problem maps into a $2D$ classical
statistical mechanics system, the $2D$ fully frustrated XY model, where the
parameter $\alpha =\left( \frac{E_{x}}{E_{C}}\right) ^{\frac{1}{2}}$ plays
the role of an inverse temperature \cite{ladder}. We work in the regime $%
E_{x}\gg E_{y}$ where the ladder is well described by a CFT with central
charge $c=2$.

We are now ready to introduce the modified ladder as represented in Fig. 2.
In order to do so let us first require the compactification of the $\varphi
^{\left( a\right) }$ variables in order to recover the angular nature of the
up and down fields. In such a way the XY-vortices, causing the
Kosterlitz-Thouless transition, are recovered. Also let us indicate the
compactified phases $\varphi ^{\left( 1\right) },\varphi ^{\left( 2\right) }$
as $\varphi _{L}^{(1)}$, $\varphi _{R}^{(2)}$ \ respectively (where $L,R$
stay for left, right components). As a second step let us introduce at point 
$x=0$ a magnetic impurity which couples the up and down phases through its
interaction with the Cooper pairs of the two legs (see Fig. 2). In the limit
of strong coupling, that is in the full screening case, such an interaction
gives rise to non trivial boundary conditions for the fields \cite{noi1}: 
\begin{equation}
\varphi _{L}^{\left( 1\right) }\left( x=0\right) =\mp \varphi _{R}^{\left(
2\right) }\left( x=0\right) -\varphi _{0}.  \label{blr}
\end{equation}%
It is interesting to notice that such a condition is naturally satisfied by
the twisted field $\phi \left( z\right) $ of our TM model (see eq. (\ref{phi}%
)). Furthermore such a field describes both the left moving component $%
\varphi _{L}^{\left( 1\right) }$ and the right moving one $\varphi
_{R}^{\left( 2\right) }$, which naturally appear in a folded description of
a system with a boundary. In fact our TM results in a chiral description of
the system just described, in terms of the chiral fields $X$ and $\phi $
(see eqs. (\ref{X}), (\ref{phi})). In Section 3 and in the Appendix we give
further details on such an issue \cite{noi1}\cite{noi2}. In particular we
adopt the $m$-reduction technique \cite{VM} which accounts for these non
trivial boundary conditions \cite{noi1}\cite{noi2} for the Josephson ladder
due to the presence of a topological defect. Furthermore its realization on
closed geometries could be relevant for the description of JJAs with non
trivial topologies, which are believed to provide a physical implementation
of an ideal quantum computer \cite{ioffe} because of the topological ground
state degeneracy which appears to be \textquotedblleft
protected\textquotedblright\ from external perturbations \cite{wen}\cite%
{noi3}.

\begin{figure}[tbp]
\centering\includegraphics*[width=0.7\linewidth]{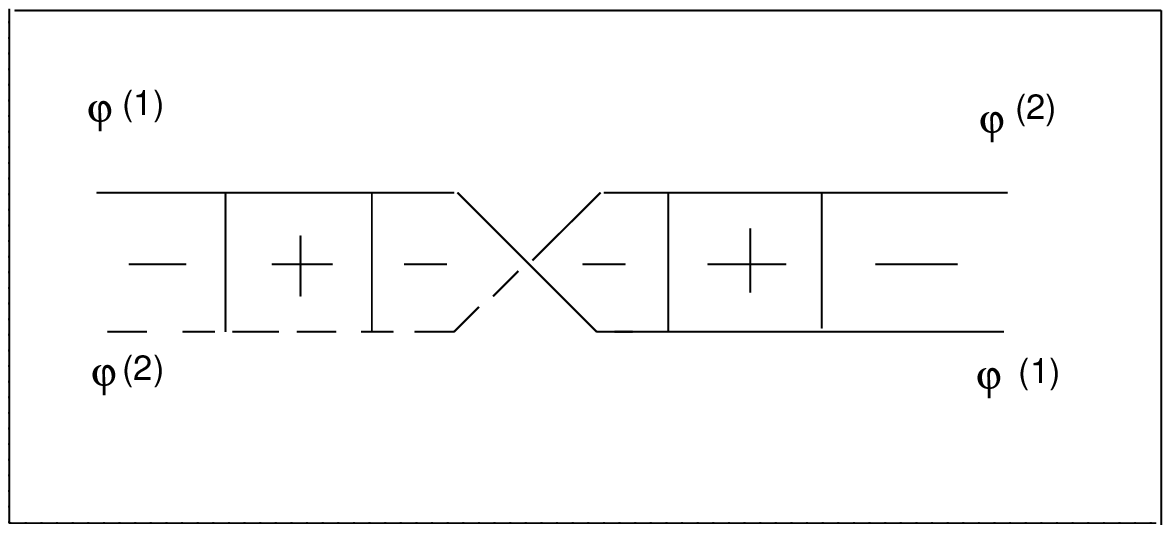}
\caption{JJL with an impurity }
\label{figura2}
\end{figure}

In a forthcoming paper \cite{noi} we will be also studying in detail two
dimensional systems with frustration, the fully frustrated XY model and a
two dimensional array of Josephson junctions in an external magnetic field
with half flux quantum per cell. Such frustrated systems represent a two
dimensional generalization of the linear chain of frustrated plaquettes
considered here. Furthermore the phase diagram of such systems \cite{granato}%
\cite{blote} can be simply understood within our TM description. Recently
conformal field theory techniques have been applied as well \cite{foda}\cite%
{lee}\cite{baseilhac}; our work follows such a line.

\section{The Twisted Model}

We are now ready to show the main steps of our construction.

\begin{enumerate}
\item We first construct the bosonic theory, i.e. the TM, and show that its
energy momentum tensor fully reproduces the Hamiltonian of eq. (\ref{ha3})
for the JJL. That allows us to describe the JJL excitations in terms of the
primary fields $V_{\alpha }\left( z\right) $ given later on in this Section
and in Sec. 4 for the torus topology.

\item Then by using standard conformal field theory techniques we show that
it is possible to construct the $N-$vertices correlator for the torus
topology in $2D$ (basically by letting the edge to evolve in
\textquotedblleft time\textquotedblright\ and to interact with external
vertex operators placed at different points). Throughout this paper we will
assume that a suitable correlator is apt to describe the ground state wave
function of the JJA at $T=0$ temperature. We must notice that such an
assumption is supported by the plasma description of the system ground state
on the plane, given later on in this Section. An analysis of the symmetry
properties of its center of mass wave function (conformal blocks), which
emerge in the presence of vortices carrying half quantum of flux ($\frac{1}{2%
}\left( \frac{hc}{2e}\right) $), will be given in Section 4.
\end{enumerate}

In this Section we recall those aspects of the TM which are relevant for the
fully frustrated ($f=\frac{1}{2}$) JJL presented in the previous Section. We
focus on the $m$-reduction procedure \cite{VM} for the special $m=2$ case
(see Ref. \cite{cgm2} for the general case), since we are interested in a
system with a $Z_{2}$ symmetry. We showed in Refs. \cite{cgm4}\cite{noi1}
that such a theory well adapts to describe a system consisting of two
parallel layers of $2D$ electrons gas in a strong perpendicular magnetic
field coupled via a defect line (a topological defect or topological twist).
The two layers edges appear coupled at a contact point carrying a magnetic
impurity (twist). The bulk electrons isospin interacts with the magnetic
impurity and in the limit of strong coupling non trivial boundary
conditions, of the $Z_{2}$ type in the considered case, for the relevant
fields emerge. In this paper we choose the \textquotedblleft
bosonic\textquotedblright\ theory, which well adapts to the description of a
system with Cooper pairs of electric charge $2e$ in the presence of a
topological defect, i.e. a fully frustrated JJL. As pointed out in the
previous Section, its ground state can be viewed as a sequence of opposite
current chiralities in adjacent plaquettes, in close analogy with the
checkerboard ground state of the two dimensional JJAs \cite{teitel}. To each
of the two legs (edges) of the ladder we assigned a chirality, making a
correspondence between up-down leg and left-right chirality states. Then we
identify in the continuum the corresponding phase fields $\varphi ^{\left(
a\right) }$, each defined on the corresponding leg, with the two chiral
fields $Q^{(a)}$($a=1,2$) of our CFT with central charge $c=2$.

In order to construct the fields $Q^{(a)}$ for the TM, we start from a
bosonic CFT with $c=1$ described in terms of a scalar chiral field $Q$
compactified on a circle with radius $R^{2}=2$. It is explicitly given by: 
\begin{equation}
Q(z)=q-i\,p\,lnz+i\sum_{n\neq 0}\frac{a_{n}}{n}z^{-n}  \label{modes}
\end{equation}%
with $a_{n}$, $q$ and $p$ satisfying the commutation relations $\left[
a_{n},a_{n^{\prime }}\right] =n\delta _{n,n^{\prime }}$ and $\left[ q,p%
\right] =i$; its primary fields are the vertex operators $U^{\alpha
_{l}}\left( z\right) =:e^{i\alpha _{l}Q\left( z\right) }:$ where $\alpha
_{l}=\frac{l}{\sqrt{2}}$, $l=1,2$. It is possible to give a plasma
description through the relation $\left\vert \psi \right\vert ^{2}=e^{-\beta
H_{eff}}$ where $\psi \left( z_{1},...,z_{N}\right) =\left\langle N\alpha
_{l}\right\vert \prod\limits_{i=1}^{N}U^{\alpha _{l}}\left( z_{i}\right)
\left\vert 0\right\rangle =\prod\limits_{i<j=1}^{N}\left( z_{i}-z_{j}\right)
^{\frac{l^{2}}{2}}$ is the ground state wave function. It can be immediately
seen that $H_{eff}=-l^{2}\sum_{i<j=1}^{N}\ln \left\vert
z_{i}-z_{j}\right\vert $ and $\beta =\frac{2}{R^{2}}=1$, that is only
vorticity $v=1,2$ vortices are present in the plasma.

From such a CFT (mother theory), using the $m$-reduction procedure, which
consists in considering the subalgebra generated only by the modes in eq. (%
\ref{modes}) which are a multiple of an integer $m$, we get a $c=m$ orbifold
CFT (daughter theory, i.e. the TM). Then the fields in the mother CFT can be
organized into components which have well defined transformation properties
under the discrete $Z_{m}$ (twist) group, which is a symmetry of the TM. By
using the mapping $z\rightarrow z^{1/m}$ and by making the identifications $%
a_{nm+l}\longrightarrow \sqrt{m}a_{n+l/m}$, $q\longrightarrow \frac{1}{\sqrt{%
m}}q$ the $c=m$ CFT (daughter theory) is obtained.

Its primary fields content, for the special $m=2$ case, can be expressed in
terms of a $Z_{2}$-invariant scalar field $X(z)$, given by 
\begin{equation}
X(z)=\frac{1}{2}\left( Q^{(1)}(z)+Q^{(2)}(z)\right) ,  \label{X}
\end{equation}%
describing the continuous phase sector of the new theory, and a twisted
field 
\begin{equation}
\phi (z)=\frac{1}{2}\left( Q^{(1)}(z)-Q^{(2)}(z)\right) ,  \label{phi}
\end{equation}%
which satisfies the twisted boundary conditions $\phi (e^{i\pi }z)=-\phi (z)$
\cite{cgm2}. More explicitly such a field can be written in terms of the
left and right moving components $\varphi _{L}^{\left( 1\right) }$, $\varphi
_{R}^{\left( 2\right) }$; then the boundary conditions given in eq. (\ref%
{blr}) are fully described by the boundary conditions for $\phi $. This will
be more evident for closed geometries, i.e. for the torus case, where the
magnetic impurity gives rise to a line defect so allowing us to resort to
the folding procedure and introduce boundary states \cite{noi1}\cite{noi2}
(see Appendix for details).

Furthermore the fields in eqs. (\ref{X})-(\ref{phi}) coincide with the ones
introduced in eq. (\ref{ha3}). In fact the energy momentum tensor for such
fields given in eq. (\ref{ttt1}) fully reproduces the second quantized
Hamiltonian of eq. (\ref{ha3}) as we will see at the end of the Section. Let
us notice that the angular nature of the phase fields in our theory takes
into account also the presence of vortices, i.e. topological excitations
which cause a Kosterlitz-Thouless transition, which are responsible for the
periodicity of the phase diagram and which were not considered in the
analysis of Ref. \cite{ladder}.

The whole TM theory decomposes into a tensor product of two CFTs, a twisted
invariant one with $c=\frac{3}{2}$ and the remaining $c=\frac{1}{2}$ one
realized by a Majorana fermion in the twisted sector. In the $c=\frac{3}{2}$
sub-theory the primary fields are composite vertex operators $V\left(
z\right) =U_{X}^{\alpha _{l}}\left( z\right) \psi \left( z\right) $ or $%
V_{qh}\left( z\right) =U_{X}^{\alpha _{l}}\left( z\right) \sigma \left(
z\right) $, where 
\begin{equation}
U_{X}^{\alpha _{l}}\left( z\right) =\frac{1}{\sqrt{z}}:e^{i\alpha _{l}X(z)}:
\label{char}
\end{equation}%
is the vertex of the continuous\ sector with $\alpha _{l}=\frac{l}{2}$, $%
l=1,...,4$ for the $SU(2)$ Cooper pairing symmetry used here. The
corresponding energy-momentum tensor is: 
\begin{equation}
T_{X}\left( z\right) =-\frac{1}{2}\left( \partial X\right) ^{2}.  \label{tc1}
\end{equation}%
Regarding the other\ component, the highest weight state in the isospin
sector, it can be classified by the two chiral operators: 
\begin{equation}
\psi \left( z\right) =\frac{1}{2\sqrt{z}}\left( :e^{i\sqrt{2}\phi \left(
z\right) }:+:e^{i\sqrt{2}\phi \left( -z\right) }:\right) ,~~~~~~\overline{%
\psi }\left( z\right) =\frac{1}{2\sqrt{z}}\left( :e^{i\sqrt{2}\phi \left(
z\right) }:-:e^{i\sqrt{2}\phi \left( -z\right) }:\right) ;  \label{neu1}
\end{equation}%
which correspond to two $c=\frac{1}{2}$ Majorana fermions with Ramond
(invariant under the $Z_{2}$ twist) or Neveu-Schwartz ($Z_{2}$ twisted)
boundary conditions \cite{cgm2}\cite{cgm4} in a fermionized version of the
theory. Let us point out that the energy-momentum tensor of the Ramond part
of the isospin sector develops a cosine term: 
\begin{equation}
T_{\psi }\left( z\right) =-\frac{1}{4}\left( \partial \phi \right) ^{2}-%
\frac{1}{16z^{2}}\cos \left( 2\sqrt{2}\phi \right) .  \label{tn1}
\end{equation}%
The Ramond fields are the degrees of freedom which survive after the
tunneling and the parity symmetry, which exchanges the two Ising fermions,
is broken.

So the whole energy-momentum tensor within the $c=\frac{3}{2}$ sub-theory
is: 
\begin{equation}
T=T_{X}\left( z\right) +T_{\psi }\left( z\right) =-\frac{1}{2}\left(
\partial X\right) ^{2}-\frac{1}{4}\left( \partial \phi \right) ^{2}-\frac{1}{%
16z^{2}}\cos \left( 2\sqrt{2}\phi \right) .  \label{ttt1}
\end{equation}%
The correspondence with the Hamiltonian of eq. (\ref{ha3}) is more evident
once we observe that the isospin current $\partial \phi $ appearing above
coincides with the term $(\partial \phi -\frac{\pi }{2})$ of eq. (\ref{ha3}%
), since the $\frac{\pi }{2}$-term coming from the frustration condition,
here it appears as a zero mode, i.e. a classical mode. That is the
frustration $\frac{\pi }{2}$ (in general $\pi f$) of the ladder cells here
in the TM construction is related to the order of the twist $Z_{2}$ ($%
Z_{1/f} $ in the general case). Besides the fields appearing in eq. (\ref%
{neu1}), there are the $\sigma \left( z\right) $ fields, also called the
twist fields, which appear in the quasi-hole primary fields $V_{qh}\left(
z\right) $. Its presence is a peculiarity of the fully frustrated XY model
in which they appear at the corner where two domain walls meet \cite{foda}.
The twist fields have non local properties and decide also for the non
trivial properties of the vacuum state, which in fact can be twisted or not
in our formalism. Such a property for the vacuum is more evident for the
torus topology, where the $\sigma $-field is described by the conformal
block $\chi _{\frac{1}{16}}$ (see Appendix).

Within this framework the ground state wave function for the plane is
described as a correlator of $N_{2e}$ Cooper pairs: 
\begin{equation}
\left\langle N_{2e}\alpha \right\vert \prod\limits_{i=1}^{N_{2e}}V\left(
z_{i}\right) \left\vert 0\right\rangle =\prod\limits_{i<i^{^{\prime
}}=1}^{N_{2e}}\left( z_{i}-z_{i^{^{\prime }}}\right) Pf\left( \frac{1}{%
z_{i}-z_{i^{^{\prime }}}}\right)
\end{equation}%
where $Pf\left( \frac{1}{z_{i}-z_{i^{^{\prime }}}}\right) =\mathcal{A}\left( 
\frac{1}{z_{1}-z_{2}}\frac{1}{z_{3}-z_{4}}...\right) $ is the
antisymmetrized product over pairs of Cooper pairs, so reproducing well
known results \cite{moore}. In a similar way we also are able to evaluate
correlators of $N_{2e}$ Cooper pairs in the presence of (quasi-hole)
excitations \cite{moore}\cite{cgm2} with non Abelian statistics \cite{nayak}.

It is now interesting to notice that the charged contribution appearing in
the correlator of $N_{e}$ electrons is just: $\left\langle N_{e}\alpha
\right\vert \prod\limits_{i=1}^{N_{e}}U_{X}^{1/2}\left( z_{i}\right)
\left\vert 0\right\rangle =\prod\limits_{i<i^{^{\prime }}=1}^{N_{e}}\left(
z_{i}-z_{i^{^{\prime }}}\right) ^{1/4}$, giving rise to a vortices plasma
with $H_{eff}=-\frac{1}{4}\sum_{i<j=1}^{N}\ln \left\vert
z_{i}-z_{j}\right\vert $ at the corresponding temperature $\beta =\frac{2}{%
R_{X}^{2}}=2$, that is it describes vortices with vorticity $v=\frac{1}{2}$!

\section{Symmetry properties of the TM conformal blocks}

In Section 3 we identified our chiral fields $Q^{(a)}$ with the continuum
limit of the Josephson phase $\varphi ^{\left( a\right) }$ defined on the
two legs of the ladder respectively and considered non trivial boundary
conditions at its ends, so constructing a version in the continuum of the
discrete system. By using standard conformal field theory techniques it is
now possible to generate the torus topology, starting from the edge theory,
just defined in the previous Section. That is realized by evaluating the $N$%
-vertices correlator 
\begin{equation}
\left\langle n\right\vert V_{\alpha }\left( z_{1}\right) \ldots V_{\alpha
}\left( z_{N}\right) e^{2\pi i\tau L_{0}}\left\vert n\right\rangle ,
\end{equation}%
where $V_{\alpha }\left( z_{i}\right) $ is the generic primary field of
Section 3 representing the excitation at $z_{i}$, $L_{0}$ is the Virasoro
generator for dilatations and $\tau $ the proper time. The neutrality
condition $\sum \alpha =0$ must be satisfied and the sum over the complete
set of states $\left\vert n\right\rangle $ is indicating that a trace must
be taken. Even though for the present paper it is not necessary to go
through such a calculation, it is very illuminating for the non expert
reader to pictorially represent the above operation in terms of an edge
state (that is a primary state defined at a given $\tau $) which propagates
interacting with external fields at $z_{1}\ldots z_{N}$ and finally getting
back to itself. In such a way a $2D$ surface is generated with the torus
topology. From such a picture it is evident then how the degeneracy of the
non perturbative ground state is closely related to the number of primary
states. Furthermore, since in this paper we are interested in the
understanding of the topological properties of the system, we can consider
only the center of mass contribution in the above correlator, so neglecting
its short distances properties. To such an extent the one-point functions
are extensively reported in the following.

On the torus \cite{cgm4} the TM primary fields are described in terms of the
conformal blocks of the $Z_{2}$-invariant $c=\frac{3}{2}$ subtheory and of
the non invariant $c=\frac{1}{2}$ Ising model, so reflecting the
decomposition on the plane outlined in the previous Section. The following
characters 
\begin{align*}
\bar{\chi}_{0}(0|\tau )& =\frac{1}{2}\left( \sqrt{\frac{\theta _{3}(0|\tau )%
}{\eta (\tau )}}+\sqrt{\frac{\theta _{4}(0|\tau )}{\eta (\tau )}}\right) , \\
\bar{\chi}_{\frac{1}{2}}(0|\tau )& =\frac{1}{2}\left( \sqrt{\frac{\theta
_{3}(0|\tau )}{\eta (\tau )}}-\sqrt{\frac{\theta _{4}(0|\tau )}{\eta (\tau )}%
}\right) , \\
\bar{\chi}_{\frac{1}{16}}(0|\tau )& =\sqrt{\frac{\theta _{2}(0|\tau )}{2\eta
(\tau )}}
\end{align*}%
express the primary fields content of the Ising model with Neveu--Schwartz ($%
Z_{2}$ twisted) boundary conditions, while 
\begin{eqnarray}
\chi _{(0)}^{c=3/2}(0|w_{c}|\tau ) &=&\chi _{0}(0|\tau )K_{0}(w_{c}|\tau
)+\chi _{\frac{1}{2}}(0|\tau )K_{2}(w_{c}|\tau )\,,  \label{mr1} \\
\chi _{(1)}^{c=3/2}(0|w_{c}|\tau ) &=&\chi _{\frac{1}{16}}(0|\tau )\left(
K_{1}(w_{c}|\tau )+K_{3}(w_{c}|\tau )\right) ,  \label{mr2} \\
\chi _{(2)}^{c=3/2}(0|w_{c}|\tau ) &=&\chi _{\frac{1}{2}}(0|\tau
)K_{0}(w_{c}|\tau )+\chi _{0}(0|\tau )K_{2}(w_{c}|\tau )  \label{mr3}
\end{eqnarray}%
represent those of the $Z_{2}$-invariant $c=\frac{3}{2}$ \ CFT. They are
given in terms of a \textquotedblleft charged\textquotedblright\ $K_{\alpha
}(w_{c}|\tau )$ contribution, (see definition given below) and a
\textquotedblleft isospin\textquotedblright\ one $\chi _{\beta }(0|\tau )$,
(the conformal blocks of the Ising Model), where $w_{c}=\dfrac{1}{2\pi i}%
\,\ln z_{c}$ is the torus variable of \textquotedblleft
charged\textquotedblright\ component. Notice that the corresponding argument
of the isospin block is $w_{n}=0$ everywhere.

In order to understand the physical significance of the $c=2$ conformal
blocks in terms of the charged low energy excitations of the system, let us
evidence their electric charge (magnetic flux contents in the dual theory,
which is obtained by exchanging the compactification radius $%
R_{e}^{2}\rightarrow R_{m}^{2}$ in the charged sector of the CFT). In order
to do so let us consider the \textquotedblleft charged\textquotedblright\
sector conformal blocks appearing in eqs. (\ref{mr1}\thinspace --\thinspace %
\ref{mr3}): 
\begin{equation}
K_{2l+i}(w_{c}|\tau )=\frac{1}{\eta \left( \tau \right) }\;\Theta \left[ 
\begin{array}{c}
\frac{2l+i}{4} \\[6pt] 
0%
\end{array}%
\right] (2w_{c}|4\tau )\,,\qquad \forall \left( l,i\right) \in \left(
0,1\right) ^{2}\,.  \label{chp}
\end{equation}%
They correspond to primary fields with conformal dimensions 
\begin{equation*}
h_{2l+i}=\frac{1}{2}\,\alpha _{\left( l,i\right) }^{2}=\frac{1}{2}\left( 
\frac{2l+i}{2}+2\delta _{\left( l+i\right) ,0}\right) ^{2}
\end{equation*}%
and electric charges $2e\left( \dfrac{\alpha _{\left( l,i\right) }}{R_{X}}%
\right) $, magnetic charges in the dual theory $\dfrac{hc}{2e}\left( \alpha
_{\left( l,i\right) }R_{X}\right) $, $R_{X}=1$ being the compactification
radius. More explicitly the electric charges (magnetic charges in the dual
theory) are the following: 
\begin{equation}
\begin{array}{llll}
l=0\,,\quad & i=0\,,\quad & q_{e}=4e\,,\quad & \left( q_{m}=2\dfrac{hc}{2e}%
\right) , \\[9pt] 
l=1\,,\quad & i=0\,,\quad & q_{e}=2e\,,\quad & \left( q_{m}=\dfrac{hc}{2e}%
\right) , \\[9pt] 
l=0\,,\quad & i=1\,,\quad & q_{e}=e\,,\quad & \left( q_{m}=\dfrac{1}{2}\,%
\dfrac{hc}{2e}\right) , \\[9pt] 
l=1\,,\quad & i=1\,,\quad & q_{e}=3e\,,\quad & \left( q_{m}=\dfrac{3}{2}\,%
\dfrac{hc}{2e}\right) .%
\end{array}%
\end{equation}%
If we now turn to the whole $c=2$ theory, the characters of the twisted
sector are given by: 
\begin{eqnarray}
\chi _{(0)}^{+}(0|w_{c}|\tau ) &=&\bar{\chi}_{\frac{1}{16}}(0|\tau )\left(
\chi _{0}^{c=3/2}(0|w_{c}|\tau )+\chi _{2}^{c=3/2}(0|w_{c}|\tau )\right) = 
\notag \\
&=&\bar{\chi}_{\frac{1}{16}}\left( \chi _{0}+\chi _{\frac{1}{2}}\right)
\left( K_{0}+K_{2}\right) ,  \label{tw1} \\
\chi _{(1)}^{+}(0|w_{c}|\tau ) &=&\left( \bar{\chi}_{0}(0|\tau )+\bar{\chi}_{%
\frac{1}{2}}(0|\tau )\right) \chi _{1}^{c=3/2}(0|w_{c}|\tau )=  \notag \\
&=&\chi _{\frac{1}{16}}\left( \bar{\chi}_{0}+\bar{\chi}_{\frac{1}{2}}\right)
\left( K_{1}+K_{3}\right) ,  \label{tw2} \\
\chi _{(0)}^{-}(0|w_{c}|\tau ) &=&\bar{\chi}_{\frac{1}{16}}(0|\tau )\left(
\chi _{0}^{c=3/2}(0|w_{c}|\tau )-\chi _{2}^{c=3/2}(0|w_{c}|\tau )\right) = 
\notag \\
&=&\bar{\chi}_{\frac{1}{16}}\left( \chi _{0}-\chi _{\frac{1}{2}}\right)
\left( K_{0}-K_{2}\right) , \\
\chi _{(1)}^{-}(0|w_{c}|\tau ) &=&\left( \bar{\chi}_{0}(0|\tau )-\bar{\chi}_{%
\frac{1}{2}}(0|\tau )\right) \chi _{1}^{c=3/2}(0|w_{c}|\tau )=  \notag \\
&=&\chi _{\frac{1}{16}}\left( \bar{\chi}_{0}-\bar{\chi}_{\frac{1}{2}}\right)
\left( K_{1}+K_{3}\right) .
\end{eqnarray}

Furthermore the characters of the untwisted sector are \cite{cgm4}: 
\begin{align}
\tilde{\chi}_{(0)}^{+}(0|w_{c}|\tau )& =\bar{\chi}_{0}(0|\tau )\chi
_{(0)}^{c=3/2}(0|w_{c}|\tau )+\bar{\chi}_{\frac{1}{2}}(0|\tau )\chi
_{(2)}^{c=3/2}(0|w_{c}|\tau )=  \notag \\
& =\left( \bar{\chi}_{0}\chi _{0}+\bar{\chi}_{\frac{1}{2}}\chi _{\frac{1}{2}%
}\right) K_{0}+\left( \bar{\chi}_{0}\chi _{\frac{1}{2}}+\bar{\chi}_{\frac{1}{%
2}}\chi _{0}\right) K_{2}\,,  \label{vac1} \\
\tilde{\chi}_{(1)}^{+}(0|w_{c}|\tau )& =\bar{\chi}_{0}(0|\tau )\chi
_{(2)}^{c=3/2}(0|w_{c}|\tau )+\bar{\chi}_{\frac{1}{2}}(0|\tau )\chi
_{(0)}^{c=3/2}(0|w_{c}|\tau )=  \notag \\
& =\left( \bar{\chi}_{0}\chi _{\frac{1}{2}}+\bar{\chi}_{\frac{1}{2}}\chi
_{0}\right) K_{0}+\left( \bar{\chi}_{0}\chi _{0}+\bar{\chi}_{\frac{1}{2}%
}\chi _{\frac{1}{2}}\right) K_{2}\,,  \label{vac2} \\
\tilde{\chi}_{(0)}^{-}(0|w_{c}|\tau )& =\bar{\chi}_{0}(0|\tau )\chi
_{(0)}^{c=3/2}(0|w_{c}|\tau )-\bar{\chi}_{\frac{1}{2}}(0|\tau )\chi
_{(2)}^{c=3/2}(0|w_{c}|\tau )=  \notag \\
& =\left( \bar{\chi}_{0}\chi _{0}-\bar{\chi}_{\frac{1}{2}}\chi _{\frac{1}{2}%
}\right) K_{0}+\left( \bar{\chi}_{0}\chi _{\frac{1}{2}}-\bar{\chi}_{\frac{1}{%
2}}\chi _{0}\right) K_{2}\,,  \label{vac3} \\
\tilde{\chi}_{(1)}^{-}(0|w_{c}|\tau )& =\bar{\chi}_{0}(0|\tau )\chi
_{(2)}^{c=3/2}(0|w_{c}|\tau )-\bar{\chi}_{\frac{1}{2}}(0|\tau )\chi
_{(0)}^{c=3/2}(0|w_{c}|\tau )=  \notag \\
& =\left( \bar{\chi}_{0}\chi _{\frac{1}{2}}-\bar{\chi}_{\frac{1}{2}}\chi
_{0}\right) K_{0}+\left( \bar{\chi}_{0}\chi _{0}-\bar{\chi}_{\frac{1}{2}%
}\chi _{\frac{1}{2}}\right) K_{2}\,,  \label{vac4} \\
\tilde{\chi}_{(0)}(0|w_{c}|\tau )& =\bar{\chi}_{\frac{1}{16}}(0|\tau )\chi
_{(1)}^{c=3/2}(0|w_{c}|\tau )=\bar{\chi}_{\frac{1}{16}}\chi _{\frac{1}{16}%
}\left( K_{1}+K_{3}\right) .  \label{vac5}
\end{align}%
Such a factorization is a consequence of the parity selection rule ($m$%
-ality), which gives a gluing condition for the \textquotedblleft
charged\textquotedblright\ and \textquotedblleft isospin\textquotedblright\
excitations. The conformal blocks given above represent the collective
states of highly correlated vortices, which appear to be incompressible. In
order to show the corresponding symmetry properties it is useful to give a
pictorial description of the conformal blocks appearing in eq. (\ref{chp}).
To such an extent let us imagine to cut the torus along the $A$-cycle. The
different primary fields then can be seen as excitations which propagate
along the $B$-cycle and interact with the external Cooper pair at point $%
w_{c}$. We can now test the symmetry properties of the characters of the
theory (given above) by simply evaluating the Bohm--Aharonov phase they pick
up while a Cooper pair is taken along the closed $A$-cycle. In order to do
that, it is important to notice that the transport of the \textquotedblright
Cooper pair\textquotedblright\ from the upper (with isospin up) leg to the
down (with isospin down) leg can be realized by a translation of the
variables $w_{c}$ and $w_{n}$, which must be identical for the
\textquotedblright charged\textquotedblright\ and the \textquotedblright
isospin\textquotedblright\ sectors. In fact it turns out that the
translation with $\Delta w_{c}=\Delta w_{n}$ allows us to describe, for
example in the twisted sector, the charge transport from leg $1$ (isospin
up) to leg $2$ (isopsin down) through the crossing point shown in Fig. 2.

So under a $2\pi $-rotation the torus variables transform as $\Delta
w_{c}=\Delta w_{n}=1$ and it is easy to check that: 
\begin{equation}
K_{0,2}(w_{c}+1|\tau )=K_{0,2}(w_{c}|\tau )\,,\qquad K_{1,3}(w_{c}+1|\tau
)=-K_{1,3}(w_{c}|\tau )\,.  \label{c18}
\end{equation}%
Let us observe that the change in sign in the last relation of eq. (\ref{c18}%
) is strictly related to the presence in the spectrum of excitations
carrying fractionalized charge quanta. Now, turning on also the isospin
sector contribution in the Cooper pair transport along the $A$-cycle, we
obtain in a straightforward way: 
\begin{equation}
\chi _{0,\frac{1}{2}}(1|\tau )=\chi _{0,\frac{1}{2}}(0|\tau )\,,\qquad \chi
_{\frac{1}{16}}(1|\tau )=i\chi _{\frac{1}{16}}(0|\tau )  \label{c19}
\end{equation}%
and the same is true for the characters $\bar{\chi}_{\beta }$. Notice that
the phase factor $i=e^{i\pi /2}$ appearing above in the transport of the
isospin "cloud" by the $\chi _{\frac{1}{16}}$ character is again due to the
presence of a half-flux.

As a result the ground state described by eq. (\ref{vac5}): 
\begin{equation}
\tilde{\chi}_{(0)}(0|w_{c}|\tau )=\bar{\chi}_{\frac{1}{16}}\chi _{\frac{1}{16%
}}\left( K_{1}+K_{3}\right)   \label{pt2}
\end{equation}%
does not change sign under the transport of a Cooper pair along the closed $A
$-cycle by the amount $\Delta w_{c}=\Delta w_{n}=1$. In fact the negative
sign coming from the continuous phase sector is compensated by the negative
sign coming from the other sector! Of course the same is true for all the
other characters of the untwisted sector, i.e. we cannot trap a half flux
quantum in the hole in the untwisted sector.

Instead in the twisted sector the ground state wave-functions show a non
trivial behavior. In fact under $\Delta w_{c}=\Delta w_{n}=1$%
\begin{equation*}
\begin{array}{l}
\chi _{\left( 0\right) }^{\pm }(1|w_{c}+1|\tau )=+i\chi _{\left( 0\right)
}^{\pm }(0|w_{c}|\tau )\,, \\[6pt]
\chi _{\left( 1\right) }^{\pm }(1|w_{c}+1|\tau )=-i\chi _{\left( 1\right)
}^{\pm }(0|w_{c}|\tau )\,.%
\end{array}%
\end{equation*}%
The change in phase given above evidences the presence of a half flux
quantum in the hole as it will be clear below. In fact in the twisted case
geometry (see Fig.\thinspace 2) the Cooper pair flows along the ladder and
changes isospin in a $2\pi $-period, so implying that in such a case the
transport of a Cooper pair from a given point $w$ on the $A$-cycle to the
same point has a $4\pi $-period, that is it corresponds to $\Delta
w_{c}=\Delta w_{n}=2$. Under this transformation the characters given above
get the following non trivial Bohm-Aharonov phase: 
\begin{equation}
\chi _{\left( 0,1\right) }^{\pm }(2|w_{c}+2|\tau )=-\chi _{\left( 0,1\right)
}^{\pm }(0|w_{c}|\tau )\,,  \label{half-flux-trapped}
\end{equation}%
so explicitly evidencing the trapping of $\frac{1}{2}\left( \frac{hc}{2e}%
\right) $ in the hole.

It is worthwhile to notice that the properties just discussed are
independent of the short distance properties of the vortices plasma, the
only crucial requirement for its stability being the neutrality condition.

\section{Brief summary with comments}

In this paper we presented a simple collective description of a ladder of
Josephson junctions with a macroscopic half flux quanta trapped in the hole.
It was shown how the phenomenon of flux fractionalization takes place within
the context of a $2D$ conformal field theory with a $Z_{2}$ twist, the TM.
The presence of a $Z_{2}$ symmetry indeed accounts for more general boundary
conditions for the fields describing the Cooper pairs propagating on the
ladder legs, which arise from the presence of a magnetic impurity strongly
coupled with the Josephson phases. For closed geometries and in the limit of
the continuum the phase fields $\varphi ^{(a)}$ defined on the two legs were
identified with the two chiral Fubini fields $Q^{(a)}$ of our TM, and a
correspondence between the energy momentum density tensor for such fields
(or better the $X$ and $\phi $ fields of eqs. (\ref{X})-(\ref{phi})) and the
Hamiltonian of eq. (\ref{ha3}) traced. For such geometries it was also
indicated that the Kosterlitz-Thouless vortices were recovered.

Furthermore it was shown that for closed geometries the JJL with an impurity
gives rise to a line defect, which can be turned into a boundary state after
employing the folding procedure. That enabled us to derive the low energy
charged excitations of the system as provided by our description, with the
superconducting phase characterized by condensation of $4e$ charges and
gapped $2e$ excitations. Finally, by simply evaluating a Bohm-Aharonov
phase, it has been evidenced that non trivial symmetry properties for the
conformal blocks emerge due to the presence in the spectrum of
fractionalized flux quanta $\frac{1}{2}\left( \frac{hc}{2e}\right) $. As it
has been explained before, that signals the presence of a topological defect
in the twisted sector of the TM. The question of an emerging topological
order in the ground state together with the possibility of providing
protected states for the implementation of a solid state qubit has been
addressed elsewhere \cite{noi3}\cite{noi}. Notice also that the different
behavior of the $2e$ and $4e$ excitations is well evidenced by the
Bohm-Aharonov phase. Indeed while the transport of a $2e\ $along the cycle
induces a -1 phase factor, in the $4e$ excitation transport the phase factor
is trivial \cite{ioffe}. This is the consequence of the symmetry of the $4e$%
\ with respect to the leg index.

It is interesting to notice that the presence of a topological defect has
been experimentally evidenced very recently for a two layers quantum Hall
system, by measuring the conduction properties between two edge states of
the system \cite{deviatov}.

We conclude by observing that it would be useful to extend our approach to a
generic frustration $f=\frac{1}{m}$.

\section{Appendix: TM boundary states}

Let us now recall briefly the TM boundary states (BS) recently constructed
in \cite{noi1}.

For closed geometries, that is for the torus, the JJL with an impurity gives
rise to a line defect in the bulk. In order to describe it we resort to the
folding procedure. Such a procedure is used in the literature to map a
problem with a defect line (as a bulk property) into a boundary one, where
the defect line appears as a boundary state of a theory which is not anymore
chiral and its fields are defined in a reduced region which is one half of
the original one. Our approach, the TM, is a chiral description of that,
where the chiral $\phi $\ field defined in ($-L/2$, $L/2)$ describes both
the left moving component and the right moving one defined in ($-L/2$, $\ 0$%
), ($0$, $L/2$) respectively, in the folded description \cite{noi1}\cite%
{noi2}. Furthermore to make a connection with the TM we consider more
general gluing conditions: 
\begin{equation*}
\phi _{L}(x=0)=\mp \phi _{R}(x=0)-\varphi _{0}
\end{equation*}
the $-$($+$) sign staying for the twisted (untwisted) sector. We are then
allowed to use the boundary states given in \cite{Affleck} for the $c=1$
orbifold at the Ising$^{2}$ radius. The $X$ field, which is even under the
folding procedure, does not suffer any change in boundary conditions \cite%
{noi1}.

The most convenient representation of such BS is the one in which they
appear as a product of Ising and $c=\frac{3}{2}$ BS. These last ones are
given in terms of the BS $|\alpha >$ for the charged boson and the Ising
ones $|f>$, $|\uparrow >$, $|\downarrow >$, according to (see ref.\cite{cft}
for details): 
\begin{align}
|\chi _{(0)}^{c=3/2}& >=|0>\otimes |\uparrow >+|2>\otimes |\downarrow > \\
|\chi _{(1)}^{c=3/2}& >=\frac{1}{2^{1/4}}\left( |1>+|3>\right) \otimes |f> \\
|\chi _{(2)}^{c=3/2}& >=|0>\otimes |\downarrow >+|2>\otimes |\uparrow >.
\end{align}%
Such a factorization naturally arises already for the TM characters \cite%
{cgm4}.

The vacuum state for the TM model corresponds to the $\tilde{\chi}_{(0)}$
character which is the product of the vacuum state for the $c=\frac{3}{2}$
sub-theory and that of the Ising one. From eqs. (\ref{vac1},\ref{vac3}) we
can see that the lowest energy state appears in two characters, so a linear
combination of them must be taken in order to define a unique vacuum state.
The correct BS in the untwisted sector are: 
\begin{align}
|\tilde{\chi}_{((0,0),0)}& >=\frac{1}{\sqrt{2}}\left( |\tilde{\chi}%
_{(0)}^{+}>+|\tilde{\chi}_{(0)}^{-}>\right) =\sqrt{2}(|0>\otimes |\uparrow 
\bar{\uparrow}>+|2>\otimes |\downarrow \bar{\uparrow}>)  \label{boud1} \\
|\tilde{\chi}_{((0,0),1)}& >=\frac{1}{\sqrt{2}}\left( |\tilde{\chi}%
_{(0)}^{+}>-|\tilde{\chi}_{(0)}^{-}>\right) =\sqrt{2}(|0>\otimes |\downarrow 
\bar{\downarrow}>+|2>\otimes |\uparrow \bar{\downarrow}>) \\
|\tilde{\chi}_{((1,0),0)}& >=\frac{1}{\sqrt{2}}\left( |\tilde{\chi}%
_{(1)}^{+}>+|\tilde{\chi}_{(1)}^{-}>\right) =\sqrt{2}(|0>\otimes |\downarrow 
\bar{\uparrow}>+|2>\otimes |\uparrow \bar{\uparrow}>) \\
|\tilde{\chi}_{((1,0),1)}& >=\frac{1}{\sqrt{2}}\left( |\tilde{\chi}%
_{(1)}^{+}>-|\tilde{\chi}_{(1)}^{-}>\right) =\sqrt{2}(|0>\otimes |\uparrow 
\bar{\downarrow}>+|2>\otimes |\downarrow \bar{\downarrow}>) \\
|\tilde{\chi}_{(0)}(\varphi _{0})& >=\frac{1}{2^{1/4}}\left( |1>+|3>\right)
\otimes |D_{O}(\varphi _{0})>  \label{continous}
\end{align}%
where we also added the states $|\tilde{\chi}_{(0)}(\varphi _{0})>$ in which 
$|D_{O}(\varphi _{0})>$ is the continuous orbifold Dirichlet boundary state
defined in ref. \cite{Affleck}. For the special $\varphi _{0}=\pi /2$ value
one obtains: 
\begin{equation}
|\tilde{\chi}_{(0)}>=\frac{1}{2^{1/4}}\left( |1>+|3>\right) \otimes |ff>.
\label{utgs}
\end{equation}%
For the twisted sector we have: 
\begin{align}
|\chi _{(0)}^{+}>& =\left( |0>+|2>\right) \otimes (|\uparrow \bar{f}%
>+|\downarrow \bar{f}>) \\
|\chi _{(1)}^{+}>& =\frac{1}{2^{1/4}}\left( |1>+|3>\right) \otimes (|f\bar{%
\uparrow}>+|f\bar{\downarrow}>)
\end{align}%
\begin{eqnarray}
|\chi _{(0)}^{-} &>&=\left( |0>-|2>\right) \otimes (|\uparrow \bar{f}%
>-|\downarrow \bar{f}>) \\
|\chi _{(1)}^{-} &>&=\frac{1}{2^{1/4}}\left( |1>+|3>\right) \otimes (|f\bar{%
\uparrow}>-|f\bar{\downarrow}>).
\end{eqnarray}%
Now, by using as reference state $|A>$ the vacuum state given in eq.(\ref%
{boud1}), we compute the chiral partition functions $Z_{AB}$ where $|B>$ are
all the BS just obtained \cite{noi1}: 
\begin{eqnarray}
Z_{<\tilde{\chi}_{((0,0),0)}||\tilde{\chi}_{((0,0),0)}>} &=&\tilde{\chi}%
_{((0,0),0)}  \label{p1} \\
Z_{<\tilde{\chi}_{((0,0),0)}||\tilde{\chi}_{((1,0),0)}>} &=&\tilde{\chi}%
_{((1,0),0)}  \label{p2} \\
Z_{<\tilde{\chi}_{((0,0),0)}||\tilde{\chi}_{((0,0),1)}>} &=&\tilde{\chi}%
_{((0,0),1)}  \label{p3} \\
Z_{<\tilde{\chi}_{((0,0),0)}||\tilde{\chi}_{((1,0),1)}>} &=&\tilde{\chi}%
_{((1,0),1)}  \label{p4} \\
Z_{<\tilde{\chi}_{((0,0),0)}||\tilde{\chi}_{(0)}>} &=&\tilde{\chi}_{(0)}
\label{p5} \\
Z_{<\tilde{\chi}_{((0,0),0)}||\chi _{(0)}^{+}>} &=&\chi _{(0)}^{+}
\label{p6} \\
Z_{<\tilde{\chi}_{((0,0),0)}||\chi _{(1)}^{+}>} &=&\chi _{(1)}^{+}.
\label{p7}
\end{eqnarray}

So we can discuss topological order referring to the characters with the
implicit relation to the different boundary states present in the system.
Also we point out that these BS should be associated to different kinds of
linear defects compatible with conformal invariance.

\textbf{Acknowledgments} - We thank Ciro Nappi and Carlo Camerlingo for
stimulating discussions and suggestions.

\end{document}